\documentclass{article}
\usepackage{spconf,amsmath,graphicx}
\usepackage{booktabs}
\usepackage{multirow}
\usepackage{graphicx}
\usepackage{xcolor}
\usepackage{url}

\title{Efficient Dense Matching for Enhanced Gaussian Splatting using AV1 Motion Vectors}
\name{Author(s) Name(s)}
\address{Author \\Affiliation(s)}

\name{Julien Zouein, Vibhoothi Vibhoothi, Fran\c{c}ois Piti\'e, Anil Kokaram}
\address{Sigmedia Group, Dept. of Electronic and Electrical Engineering,\textit{Trinity College Dublin}, Ireland\\
\{zoueinj, vibhootv, pitief, anil.kokaram\}@tcd.ie \thanks{This work was funded by the Horizon CL4 2022, EU Project Emerald, 101119800; and YouTube \& Google Faculty Awards.}}

\begin{document}
%
\maketitle

\begin{abstract}
3D Gaussian Splatting (3DGS) has emerged as a prominent framework for real-time, photorealistic scene reconstruction, offering significant speed-ups over Neural Radiance Fields (NeRF). However, the fidelity of 3DGS representations remains heavily dependent on the quality of the initial point cloud. While standard Structure-from-Motion (SfM) pipelines using COLMAP provide adequate initialisation, they often suffer from high computational costs and sparsity in textureless regions, which degrades subsequent reconstruction accuracy and convergence speed. In this work, we introduce an AV1-based feature detection and matching pipeline that significantly reduces SfM processing overhead. By leveraging motion vectors inherent to the AV1 video codec, we bypass computationally expensive exhaustive matching while maintaining geometric robustness. Our pipeline produces substantially denser point clouds, with up to eight times as many points as classical SfM. We demonstrate that this enhanced initialisation directly improves 3DGS performance, yielding an 9-point increase in VMAF and a 63\% average reduction in training time required to reach baseline quality. The project page: \url{https://sigmedia.tv/AV1-3DGS.github.io/}

\end{abstract}

\section{Introduction}
\label{sec:intro}
The field of 3D scene reconstruction has recently undergone a paradigm shift with the introduction of 3D Gaussian Splatting (3DGS)~\cite{3dgs}. By combining the utility of explicit geometric representations with differentiable rendering, 3DGS offers real-time rendering speeds and high-fidelity view synthesis, addressing the primary computational bottlenecks of Neural Radiance Fields (NeRF~\cite{nerf}).

Despite these capabilities, 3DGS quality is intrinsically linked to the underlying geometry provided during initialisation. Standard workflows rely on sparse point clouds generated via Structure from Motion (SfM). When we are generating 3DGS from a video captured from a mobile device/camera, the point cloud can be sparse. When these initial clouds are overly sparse, which is common in textureless regions or complex indoor environments, the resulting 3DGS models frequently exhibit visual artefacts such as blurred boundaries, floaters, and geometric inconsistencies. Furthermore, the conventional approach to mitigating sparsity involves computationally intensive dense matching strategies, which hinder rapid content creation and scalability.

In this paper, we propose a framework that enhances the initialisation of 3D scenes by bridging the gap between sparse SfM and heavy dense matching. We propose generating point-correspondences for SfM by extracting motion directly from a compressed video bitstream using AV1~\cite{technical_overview_av1} encoding in commonly available hardware. By repurposing video coding metadata in this way, we can generate high-density geometry at a fraction of the computational cost of traditional methods.

The primary contributions of this work are: 
\begin{enumerate} 
\item An end-to-end SfM pipeline that leverages AV1 motion vectors to generate dense point clouds without exhaustive matching in pixel-domain.
\item Improved 3DGS rendering quality (+9 points in VMAF) by exploiting dense point clouds from our new pipeline.
\end{enumerate}




\section{Related Work}
\label{sec:related}
Structure from Motion (SfM) serves as the geometric foundation for most modern 3D reconstruction frameworks, ranging from classical photogrammetry to Neural Radiance Fields (NeRF)~\cite{nerf} and 3D Gaussian Splatting (3DGS)~\cite{3dgs}. 
While these rendering techniques have evolved to achieve real-time performance, their visual fidelity remains critically dependent on the density and accuracy of the initial point cloud. 
Consequently, the primary computational bottleneck has shifted to the pre-processing stage, the feature extraction and matching front-ends. To address this, recent research has focused on either optimising these geometric pipelines~\cite{superpoint, lightglue, glomap} or exploring alternative data sources, such as compressed video bitstreams~\cite{h264-2d-3d, mov-slam}, to accelerate scene understanding.

{\noindent \textbf{Feature Learning and Matching.}}
Traditional SfM pipelines, such as COLMAP~\cite{colmap}, typically rely on hand-crafted features like SIFT~\cite{SIFT}.
While robust, these are computationally intensive to extract and match exhaustively. 
Deep learning offers powerful alternatives for both extraction (e.g., SuperPoint~\cite{superpoint}, DISK~\cite{disk}) and matching (e.g., SuperGlue~\cite{superglue}, LightGlue~\cite{lightglue}). While LoFTR~\cite{loftr}, an end-to-end model, bypasses detection entirely (jointly performs feature extraction and the matching process).
Although these learned methods provide superior invariance to viewpoint changes, they incur significant GPU costs. 
Similarly, while recent global SfM solvers like GLOMAP~\cite{glomap} improve the scalability of the reconstruction back-end, they still depend on these expensive front-ends. 
Regardless of whether the solver is incremental or global, the computational burden of correspondence generation remains a limiting factor, motivating the search for more efficient proxies.

{\noindent \textbf{Improving Dense 3D Reconstruction.}} 
While NeRF~\cite{nerf} and Instant-NGP~\cite{instant-ngp} made significant improvements for view synthesis, 3DGS~\cite{3dgs} has emerged as the dominant approach for real-time, high-fidelity rendering. However, 3DGS is sensitive to its initialisation state. Sparse point clouds frequently result in visual artefacts such as floaters and blurred geometry. Although methods exist to improve the quality of point clouds specifically for 3DGS~\cite{eap-gs}, they typically operate as a post-processing step, adding further computational overhead. 
Our approach addresses this by improving the density of the SfM output itself, ensuring robust initialisation without additional processing.


{\noindent \textbf{Using Compression Data for Visual Tasks:} }
To bypass the cost of pixel-domain processing, researchers have increasingly utilised metadata from video codecs. 
In 2010, Pourazad et al.~\cite{h264-2d-3d} used H.264 motion vectors for depth estimation in 2D-to-3D conversion. 
More recently, in 2023, MoV-SLAM~\cite{mov-slam} demonstrated that hardware-encoded H.264 motion vectors could drive a real-time visual SLAM system on low-power devices (30 FPS on a Raspberry Pi 4), outperforming ORB-SLAM3~\cite{orb_slam_3} in specific contexts. But in both of these cases, the image data is still used to validate candidate matches extracted from compressed domain motion. That implies that the cost of matching is still high. Zouein \textit{et al}. presented a new matching process~\cite{icir_2025_jz}, relying on Motion Vectors extracted from AV1 to generate keypoints matches, without having to process pixels. This therefore reduces computation and processing time. We discuss this in the next section.

\section{3D Point-Cloud Generation with AV1}
\label{sec:methodology}
Extending the work initiated by Zouein \textit{et al.}~\cite{icir_2025_jz}, we employ the AV1-based matching pipeline and combine it with COLMAP incremental mapping to generate point clouds. 

\textbf{Matching:} Zouein \textit{et al.} in their work~\cite{icir_2025_jz} use AV1 video codec. Because they are using AV1, they upsample the motion field (using zero order hold) from non-uniform blocks to yield a motion vector for every $4 \times 4$ blocks. Matches that are persistent across many frames are more likely to be useful for point cloud reconstruction, and so they build motion trajectories by accumulating motion vectors themselves. They terminate a trajectory when the cosine difference between motion-compensated vectors is greater than some threshold, since that indicates loss of track. The resulting matches are dense, and they retain matches only if they persist across more than 3 frames along a trajectory. The reason why this is computationally efficient is not only that there is no need to use image data itself, but also that hardware AV1 encoding can be used to generate point correspondences in real time, even with very high resolution footage.

\textbf{3D Gaussian Splats:} As discussed, in this work we generate 3D point clouds by incorporating the matches into COLMAP~\cite{colmap} incremental mapping process. This results in the camera path through our video sequence as well as a single point cloud in 3D space. That single point cloud is presented to the official 3DGS implementation\footnote{https://github.com/graphdeco-inria/gaussian-splatting} including the camera geometry. We train the Gaussian models using 50\% of our input frames, and use this to synthesise frames at the same location of the remaining observed frames.



\textbf{Assessing Quality:} We place greater emphasis on quality evaluation in 3D space and the final synthesised images. 3D point cloud quality evaluation ideally requires some kind of ground truth point cloud. This is not available for the real video sequences we use in this work. To mitigate this problem somewhat, we use the output of the classical pipeline with Exhaustive matching as ``pseudo-ground truth''. This is the default in Gaussian Splatting. Clearly, this introduces a problem when the pseudo-ground truth actually does not perform well. Therefore, comparison of synthesised images provides the ultimate evaluation metric. Hence we synthesised pictures at timestamps for which we already have ground truth images, and as discussed before, we do not use those locations in training the 3DGS technique.

To more completely assess the impact of competing feature and matching combinations we assess our AV1-based initialisation against four configurations: (1) SIFT~\cite{SIFT} features with \textit{Exhaustive} matching; (2) SIFT features with \textit{Sequential} matching; (3) \textit{DISK}~\cite{disk} features with \textit{LightGlue}~\cite{lightglue} (DISK) and (4) \textit{SuperPoint}~\cite{superpoint} features with \textit{SuperGlue}~\cite{superglue} (SP). 

Finally, we note that while dense optical flow and KLT front-ends exist, they require pixel-domain estimation at inference time. In contrast, this approach reuses motion already computed by the encoder; we therefore restrict our comparison to standard SfM feature/matcher families.
    
\section{Experimental Setup}
\label{sec:experiments}

Our evaluation focuses on the downstream impact of these denser point clouds on 3DGS performance. Hence we compare output from proposed AV1-based pipeline against state-of-the-art SfM techniques.

The SfM pipeline (COLMAP) was executed on a 12th Gen Intel Core i7-12700K CPU and 64GB RAM. Deep learning-based feature extraction (used in Disk, Superpoint) used an NVIDIA RTX A4000 GPU running Ubuntu~22.04. Hardware-accelerated encoding was performed on a NVIDIA RTX~6000 Ada, whilst the 3DGS training was conducted on an NVIDIA RTX PRO 6000 Blackwell. Following\cite{pcs_2025_jz} which validated Motion Vectors quality, we use hardware AV1 encoding (NVENC-AV1, Ada Lovelace) via FFmpeg \texttt{n6.0-22}, using single intra-frame and backward reference frames only (S3-SCC-03~\cite{av1_stream_conf, aomctc}).\footnote{FFmpeg command: \texttt{ffmpeg -i \$input\_vid -c:v av1\_nvenc -tune hq -preset p1 -rc constqp -qp qp -g 9999 -b\_ref\_mode 0 -bf 0 output.ivf}}

For Gaussian Splatting, we use the official implementation (Kerbl et al.~\cite{2023_gsplat_orig_paper}). As our dataset comprises primarily large outdoor scenes, we used the flags~\texttt{\small --position\_lr \_init} and \texttt{\small --position\_lr\_final} to optimise both training and rendering. To ensure a rigorous testing process, we enforce the 50/50 train-test split described in Section~\ref{sec:methodology}.

\subsection{Dataset}
Our dataset comprises seven video sequences categorised by their source and content characteristics. To represent complex urban environments with rapid ego-motion, we use two driving sequences from the KITTI Odometry dataset~\cite{kitti_odometry}: the initial 231 frames of Seq.~00 (1241$\times$376) and the first 131 frames of Seq.~10 (1226$\times$370). 
To incorporate handheld urban scenes, we include three sequences captured using an iPhone 15 Pro at 10fps: Paris Seq.~1 (117 frames, 1080$\times$1920), Paris Seq.~2 (122 frames, 1920$\times$1080), and Dublin Seq.~1 (329 frames, 1080$\times$1920). 
Finally, to analyse the reconstruction of fine details, we introduce two high-resolution sequences (2160$\times$3840, 10fps) captured specifically for this study: \textit{Boston Seq.~1} (55 frames), which captures a bas-relief exhibiting intricate high-frequency geometry, and \textit{Nature} (67 frames), which focuses on complex organic textures such as foliage.

\subsection{Evaluation Metrics}
To assess point cloud fidelity we use reprojection error and report two other metrics discussed more clearly in the supplementary material (Hausdorff and Chamfer distances).

To assess the quality of the rendered images using Gaussian Splatting, we employ a suite of classical and perceptual picture quality metrics. We report PSNR and SSIM as standard measures of fidelity. To specifically address video quality, we also report VMAF~\cite{vmaf_paper}, an important perceptual quality metric widely adopted in the streaming video industry. We also include LPIPS~\cite{lpips}, a metric based on deep network activation functions that correlates well with human perceptual similarity; for LPIPS, lower scores indicate better performance. None of these actually measures sharpness which is an important indicator of the ability of 3D reconstruction to reproduce texture. Therefore we include the Q-metric~\cite{2009_q_metric_paper_qomex}, a non-reference sharpness metric with good correlation with perceptual sharpness.

\section{Results and Discussions}
\label{subsec:result_and_analysis}

\begin{figure}[t!]
    \centering
    \begin{tabular}{c}
      \includegraphics[width=1\linewidth]{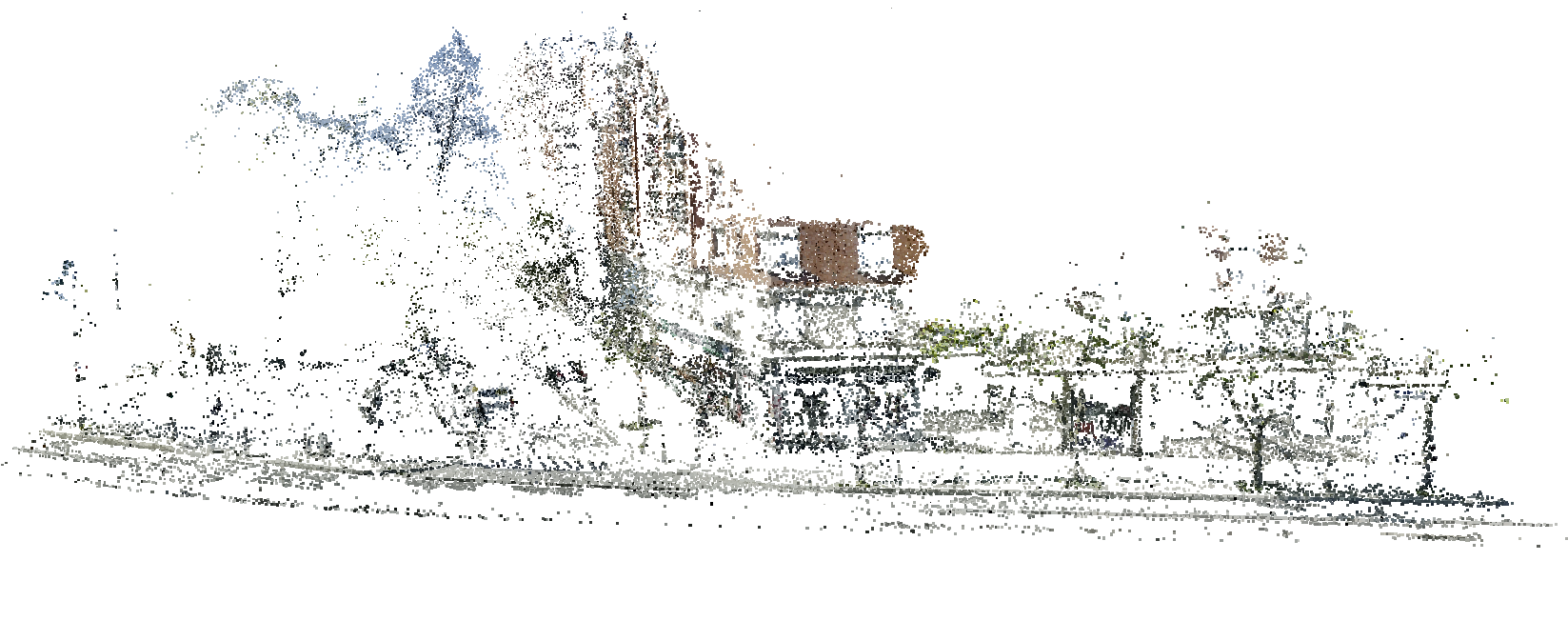} \label{paris_baseline}\\
      \includegraphics[width=1\linewidth]{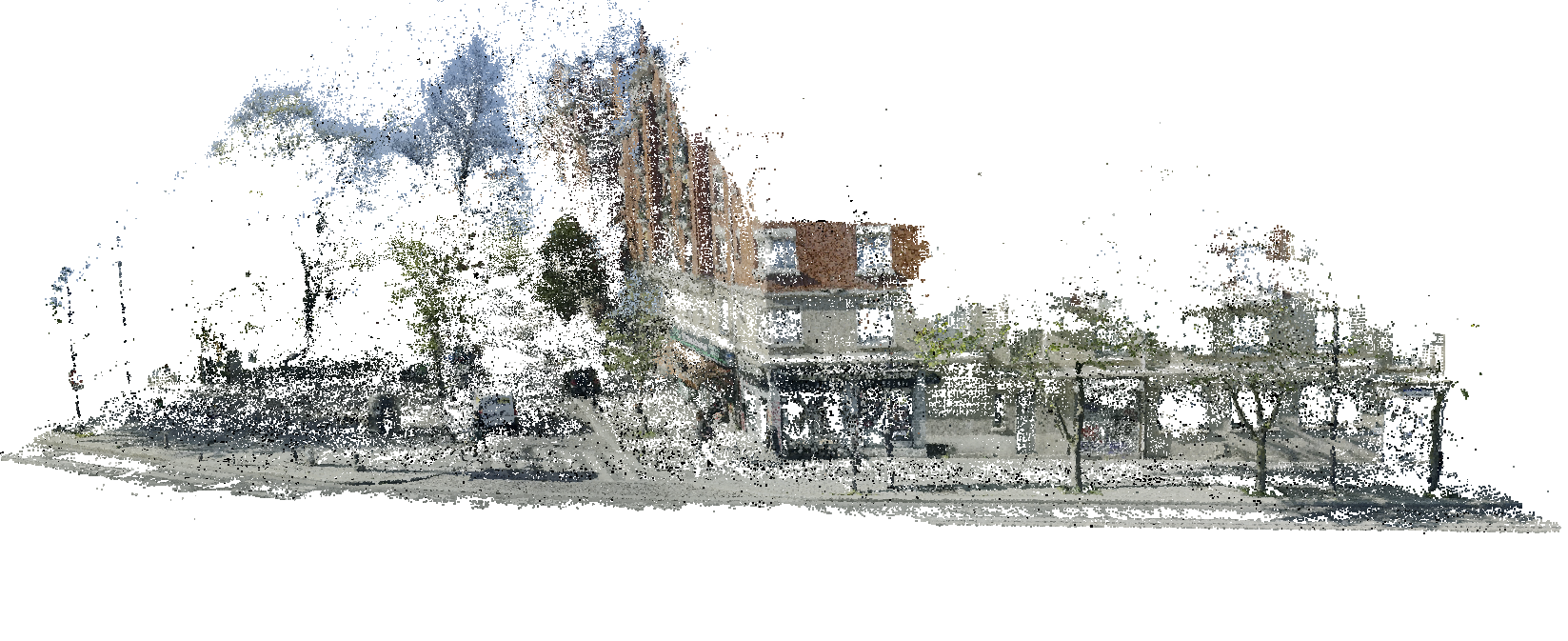}\label{paris_av1} \\
    \end{tabular}
    \caption{
    Visual comparison of reconstructed 3D point clouds for Paris Seq.~1. Top: Baseline COLMAP reconstruction using SIFT features with Exhaustive matching (48k points). Bottom: Our proposed AV1-based pipeline (621k points). While the overall scene structure remains consistent, our method produces a significantly denser point cloud, effectively capturing finer geometric details.}
    \label{3d_point_comparison}
    \vspace{-1em}
\end{figure}

Figure~\ref{3d_point_comparison} shows point clouds generated from our pipeline (bottom) compared to the baseline (top) for Paris Seq. 1. See supplementary material for further examples. Clearly our point cloud is denser (621k versus 48k points) and captures more of the fine quality texture information. For this particular sequence, the mean reprojection error was 0.51 compared with 0.56 for baseline. This observation is also supported by Hausdorff and Chamfer distances. However, as discussed previously, this favourable comparison is biased by our denser point cloud. Much better evidence is contained in our picture quality results discussed next.

\begin{table*}
\centering
\caption{Comparison of Structure-from-Motion (SfM) methods across reconstruction iterations for 3D Gaussian Splatting (3DGS). Metrics represent the mean performance across all seven test sequences. The proposed AV1-SfM method consistently outperforms traditional and learning-based baselines across all metrics. We achieve higher quality scores at 30k iterations than most baselines achieve at 42k, demonstrating improved convergence efficiency.}
\label{tab:sfm_comparison}
\resizebox{\textwidth}{!}{%
\begin{tabular}{@{}llcccccc@{}}
\toprule
\textbf{} & \textbf{Method} & \textbf{Iterations} & \textbf{PSNR-Y (dB) $\uparrow$} & \textbf{SSIM (dB) $\uparrow$} & \textbf{LPIPS $\downarrow$} & \textbf{VMAF $\uparrow$} & \textbf{$\Delta$ Q $\downarrow$} \\ \midrule
\multirow{5}{*}{Average} 
 & Exhaustive & 30k / 42k & 26.22 / 26.31 & 8.36 / 8.44 & 0.256 / 0.251 & 69.91 / 71.56 & 1.89 / 1.75 \\
 & Sequential & 30k / 42k & 27.03 / 27.14 & 8.81 / 8.89 & 0.236 / 0.229 & 74.77 / 75.75 & 1.10 / 1.00 \\
 & SuperPoint (SP)~\cite{superpoint} & 30k / 42k & 25.11 / 25.25 & 7.68 / 7.77 & 0.296 / 0.289 & 62.94 / 64.23 & 2.30 / 2.09 \\ 
 & Disk~\cite{disk} & 30k / 42k & 27.09 / 27.19 & 8.81 / 8.91 & 0.238 / 0.230 & 75.72 / 76.76 & 1.01 / 0.90 \\
 & AV1-SfM (Ours) & 30k / 42k & \textbf{27.48 / 27.51} & \textbf{9.24 / 9.27} & \textbf{0.198 / 0.195} & \textbf{78.26 / 78.86} & \textbf{0.68 / 0.66} \\
\bottomrule
\end{tabular}%
}

\end{table*}

Table~\ref{tab:sfm_comparison} summarises picture quality performance across all seven video sequences at two critical training milestones: 30,000 (30k) and 42,000 (42k) iterations. These results demonstrate that the proposed AV1-SfM method consistently outperforms both classical and learning-based baselines across different quality metrics.
Our method achieves a PSNR of 27.51 dB at the 42k iteration, representing an improved quality metric score compared to learning-based feature matchers such as Disk~\cite{disk} (27.19 dB) and SuperPoint~\cite{superpoint} (25.25 dB). 
This performance gap is much more pronounced in perceptual metrics; AV1-SfM yields an LPIPS score of 0.195, a notable improvement over the 0.289 observed with SuperPoint and the 0.256 with the baseline.

\begin{figure}[t!]
    \centering
    \begin{tabular}{cc}
      \includegraphics[width=0.44\linewidth]{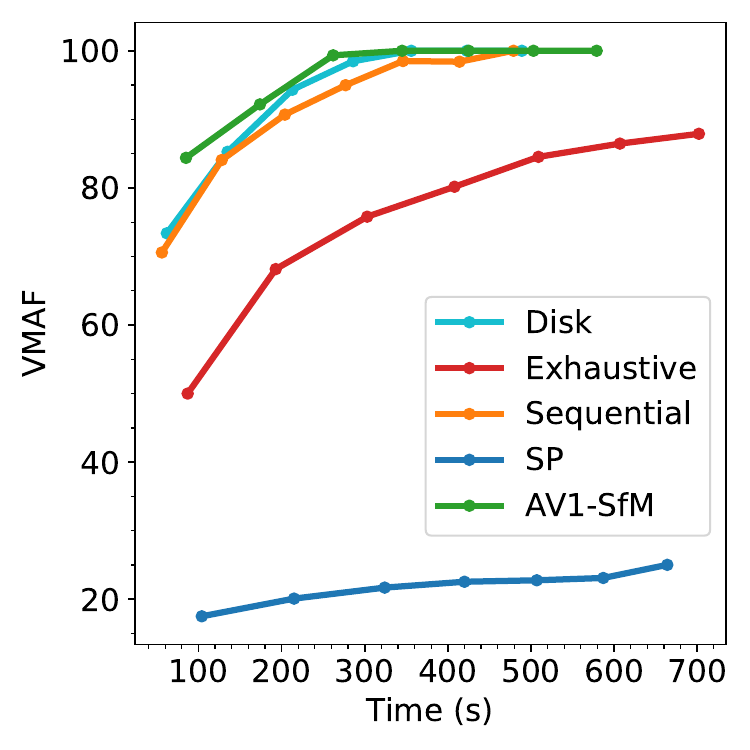}\label{kitti-train-vmaf}   & 
      \includegraphics[width=0.44\linewidth]{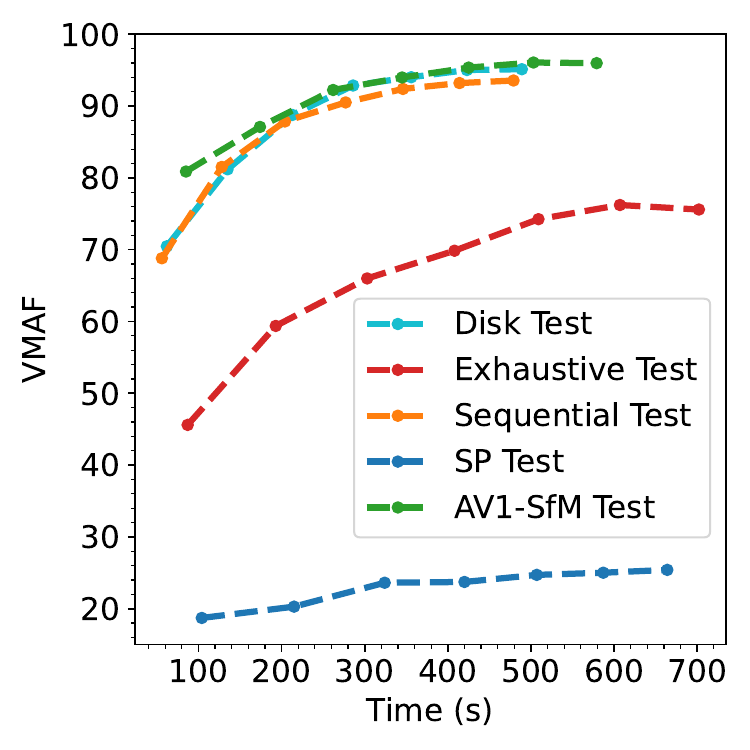} \label{kitti-test-vmaf}\\
      \includegraphics[width=0.44\linewidth]{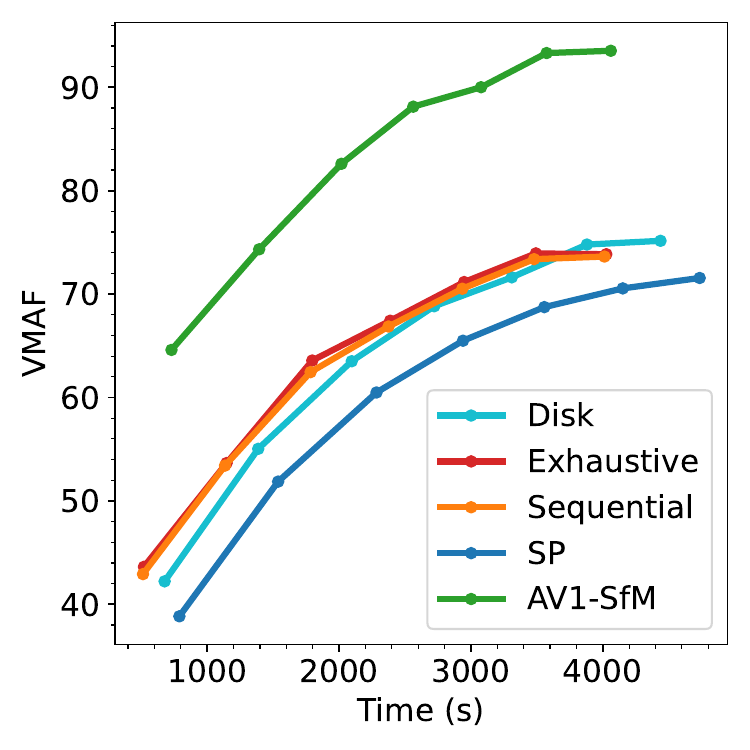} &
      \includegraphics[width=0.44\linewidth]{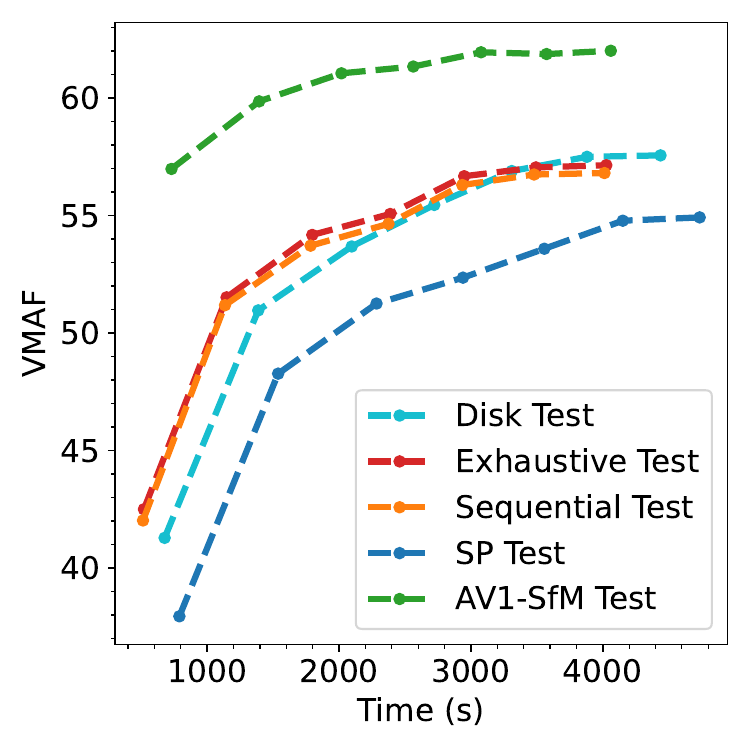}
    \end{tabular}
    \caption{Evolution of VMAF value during training. Top row: Train (left)-Test (right) results with Kitti Seq 00, Bottom row: Same for Nature Test Sequence. Our technique reaches higher quality in less training time than other matching techniques.}
    \label{VMAF Training}
    \vspace{-1.5em}
\end{figure}

{\noindent \textbf{Convergence and Geometric Debt.}}
A primary contribution of this work is the marked improvement in convergence efficiency, as visualised in the evolution of VMAF (Figure~\ref{VMAF Training}). In both urban Kitti (Figure~\ref{VMAF Training} top row) and organic Nature (Figure~\ref{VMAF Training} bottom row) sequences, the AV1-SfM curves (green) exhibit a higher initial state compared to all other methods.
This ``head start'' is attributed to the reduction of what we term ``geometric debt''. In standard 3DGS workflows, the optimiser spends thousands of iterations generating new Gaussians to fill the structural voids left by sparse SfM. By providing high-density initialisation at the start, AV1-SfM allows the GPU to focus its computational budget on refining colour and covariance parameters immediately. This results in a 63\% reduction in the total training time required to reach baseline quality (averaged over all sequences).

{\noindent \textbf{Impact on Perceptual Metrics.}}
By analysing Figure~\ref{VMAF Training}, we observe an improvement on perceptual quality. In the Nature sequence, characterised by complex, fine-grained foliage, the VMAF gap (Figure~\ref{VMAF Training}c-d) is notably wide. The disparity between Kitti and Nature test /sequences suggests that while methods like DISK and Sequential matching may reach comparable VMAF for kitti, they fail to achieve the same level of perceptual quality for complex scenes. Specifically, our method reaches a VMAF score within the first 1,000 seconds of training that the baseline exhaustive matching cannot achieve even after a full 4,000-second cycle. The superior VMAF and SSIM scores suggest that our framework is particularly adept at preserving high-frequency textural details. 

{\noindent \textbf{Geometric Consistency using Q-Metric.}}
To evaluate structural integrity, we employ the Q-metric (lower is better), which measures the sharpness of 3DGS. This, in turn, provides diagnostic insight into geometric consistency. We measure the difference between $\hat{Q}$ using each technique with $Q_\text{ref}$ from the ground truth image i.e. $\Delta Q = |Q_\text{ref}-\hat{Q}|$ 
Our framework achieves the lowest error ($\Delta Q=0.66$ at 42k), whereas the Exhaustive matching approach yields a significantly higher error ($\Delta Q=1.75$). High Q-metric scores in traditional pipelines often stem from the low density of 3D points in repetitive areas, forcing Gaussian splats to expand incoherently to ``fill the space''. We also observe consistently high VMAF plateaus in the training process (Figure~\ref{VMAF Training}). The stability of the Q-metric, coupled with a plateauing VMAF score in training for our method, confirms that AV1 motion vectors provide a reliable geometric backbone. 


\begin{figure*}
    \centering
    \resizebox{\textwidth}{!}{%
    \begin{tabular}{cccccc}
        Exhaustive & Sequential & SP & Disk & AV1-SfM (ours) & Ground Truth \\
      \includegraphics[width=0.16\linewidth]{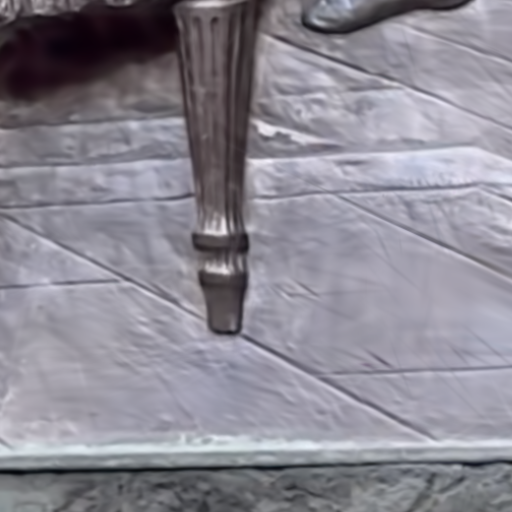} & 
      \includegraphics[width=0.16\linewidth]{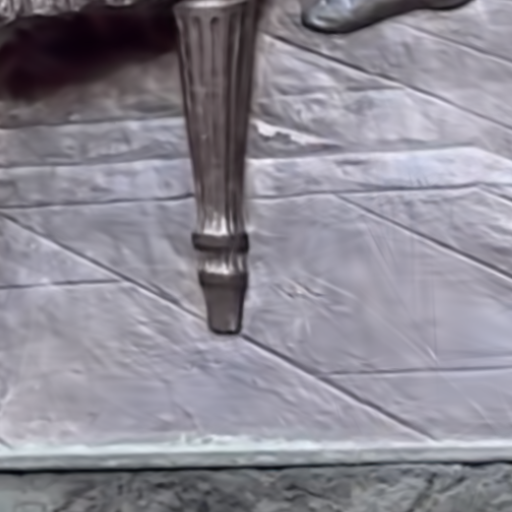} & 
      \includegraphics[width=0.16\linewidth]{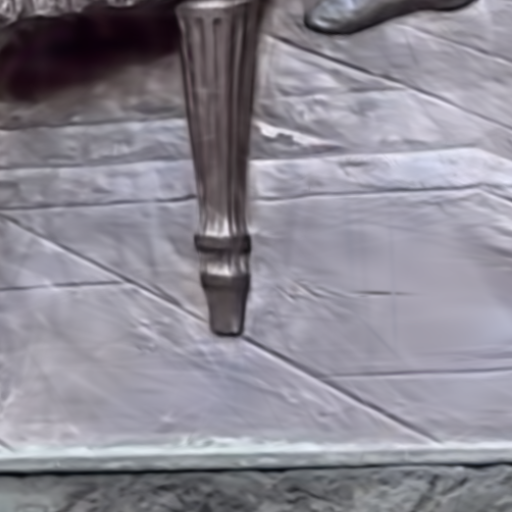} &
      \includegraphics[width=0.16\linewidth]{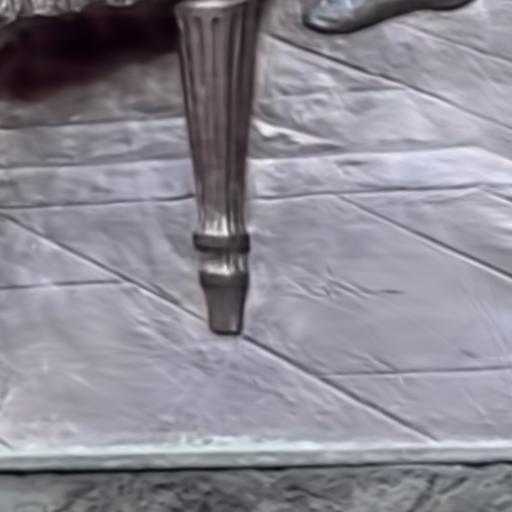} &
      \includegraphics[width=0.16\linewidth]{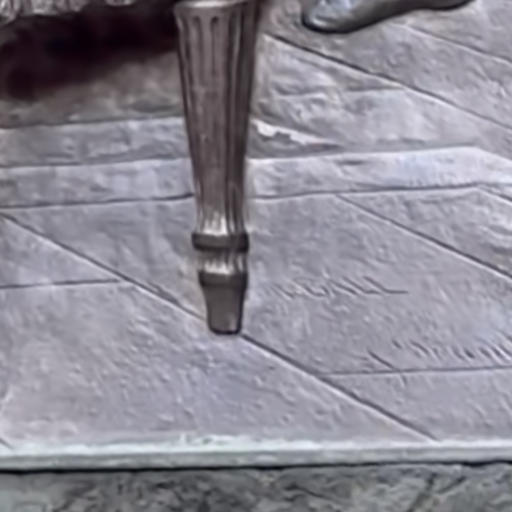} &
      \includegraphics[width=0.16\linewidth]{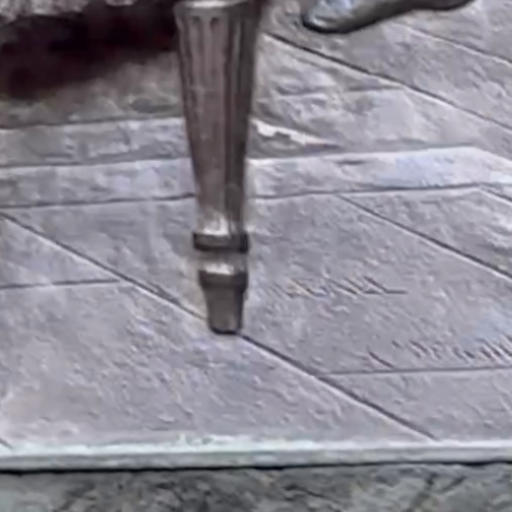}
      \\
      $\Delta$Q = 0.01 / LPIPS=0.18 & $\Delta$Q=0.11 / LPIPS=0.18 & $\Delta$Q=0.14 / LPIPS=0.25 & $\Delta$Q=0.14 / LPIPS=0.21 & $\Delta$Q=0.25 / LPIPS=0.11 & \\
      \includegraphics[width=0.16\linewidth]{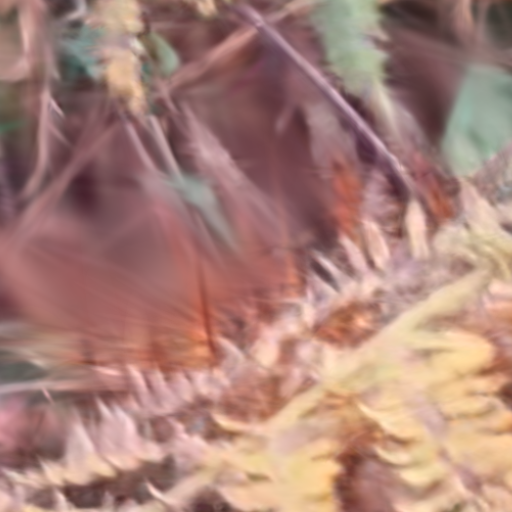} & 
      \includegraphics[width=0.16\linewidth]{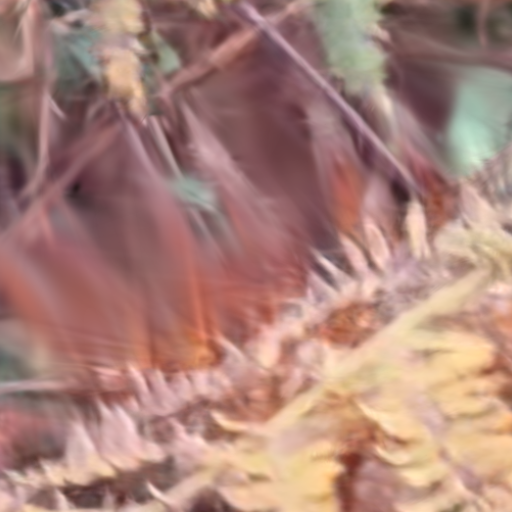} & 
      \includegraphics[width=0.16\linewidth]{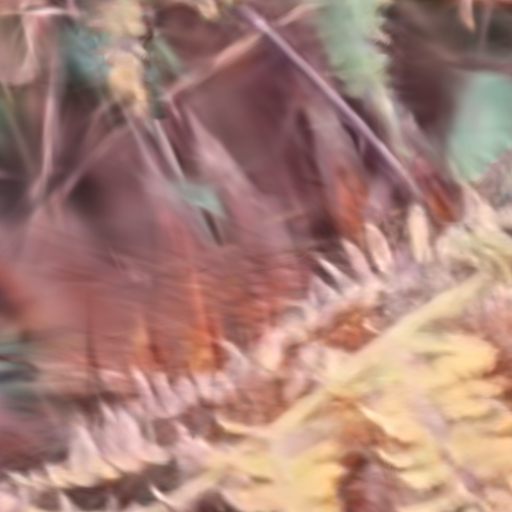} &
      \includegraphics[width=0.16\linewidth]{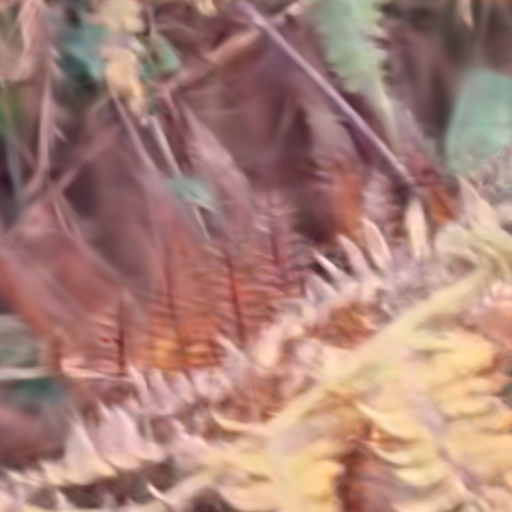} &
      \includegraphics[width=0.16\linewidth]{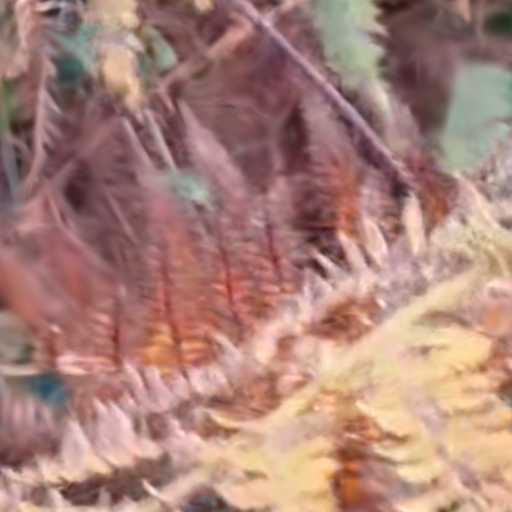} &
      \includegraphics[width=0.16\linewidth]{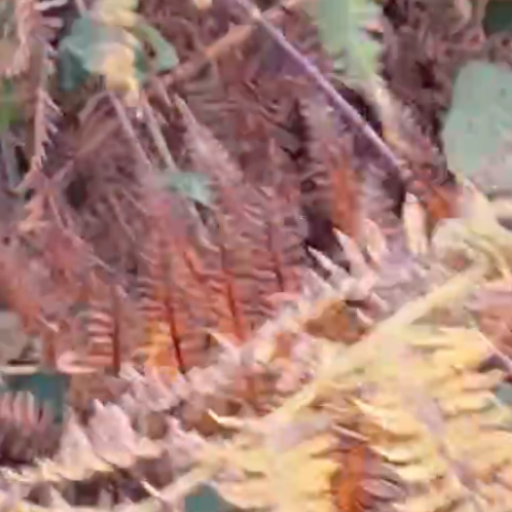}
      \\
      $\Delta$Q=1.23 / LPIPS=0.41 & $\Delta$Q=1.34 / LPIPS=0.42 & $\Delta$Q=1.35 / LPIPS=0.45 & $\Delta$Q=0.97 / LPIPS=0.39 & $\Delta$Q=1.04 / LPIPS=0.33 & \\
      \includegraphics[width=0.16\linewidth]{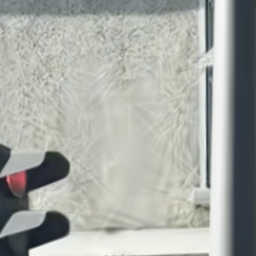} & 
      \includegraphics[width=0.16\linewidth]{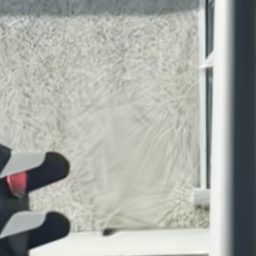} & 
      \includegraphics[width=0.16\linewidth]{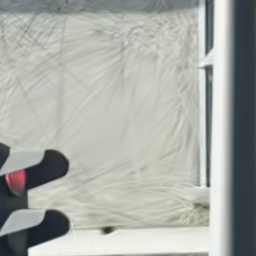} &
      \includegraphics[width=0.16\linewidth]{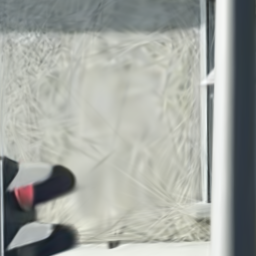} &
      \includegraphics[width=0.16\linewidth]{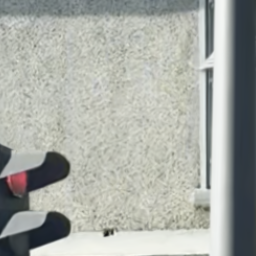} &
      \includegraphics[width=0.16\linewidth]{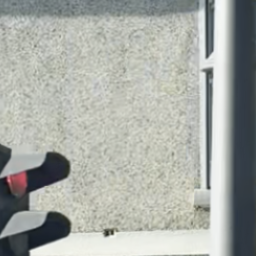}
      \\
      $\Delta$Q=0.21 / LPIPS=0.24 & $\Delta$Q=0.21 / LPIPS=0.21 & $\Delta$Q=0.58 / LPIPS=0.33 & $\Delta$Q=0.53 / LPIPS=0.42 & $\Delta$Q=0.06 / LPIPS=0.08 &
    \end{tabular}
    }
    \caption{Visualisation of a 512$\times$512 patch from sequence \textit{Boston Vid 1} (Test Set, iteration 30000), a 512$\times$512 patch from sequence \textit{Nature} (Test set, iteration 30000) and a 256$\times$256 patch from sequence \textit{Paris Seq 1} (Test Set, iteration 30000). AV1-SfM shows better details and texture reconstruction than any other technique. $\Delta$Q and LPIPS values for each patch are specified.}
    \label{patch_comparison}
    \vspace{-1em}
\end{figure*}

{\noindent \textbf{Visual Analysis.}} The visual evidence presented in Figure~\ref{patch_comparison} reinforces these findings. In the \textit{Boston Seq. 1} sequence, our method is the only one to faithfully reconstruct minute pavement inscriptions that are smoothed over by alternative techniques. Similarly, in the \textit{Nature} sequence, classic SIFT-based reconstructions fail to capture the individual structure of leaves, resulting in a generalised, blurred appearance. Furthermore, our approach demonstrates superior handling of textureless patches, such as those in \textit{Paris Seq 1}. In contrast, standard 3DGS struggles with these regions by over-extending splats (leading to a `bloated' appearance). This indicates that the dense point cloud generated via AV1 is not just a faster alternative, but a more robust one for complex, real-world environments. Our method consistently achieves the lowest LPIPS value in all patches. $\Delta$Q shows coherent results; however, it also shows that this metric cannot be used ``alone'', as it does not fully represent perceptual quality.

The analysis of the train-test split reveals a smaller `generalisation gap' for our framework. As the initialisation is geometrically accurate, the model learns the actual scene structure rather than over-fitting to training view intensities. This leads to more stable performance on unseen test frames, with fewer visual artefacts when the virtual camera moves outside the primary training trajectory. This effectively bridges the gap between sparse SfM and expensive dense matching.
\vspace{-1em}

\section{Conclusion}
\label{subsec:conclusion_and_future_work}
In this work, we introduced a new pipeline that uses AV1 motion vectors to initialise 3DGS. 
By repurposing compressed video metadata, we generate dense geometric seeds without the computational cost of exhaustive matching. 
Our experiments demonstrate that this enhanced density directly translates to superior rendering fidelity, achieving higher perceptual quality (VMAF: 78.86 v/s 71.56, Q-metric: 0.66 vs 1.75) while reducing training time by 63\% compared to standard baselines (Exhaustive matching). 
This approach effectively improves the creation of photorealistic digital twins from User Generated Content (UGC), bridging the gap between efficient video coding and immersive 3D reconstruction.
\bibliographystyle{IEEEbib}
\bibliography{references}

@inproceedings{kitti_odometry,
  author = {Andreas Geiger and Philip Lenz and Raquel Urtasun},
  title = {Are we ready for Autonomous Driving? The KITTI Vision Benchmark Suite},
  booktitle = {Conference on Computer Vision and Pattern Recognition (CVPR)},
  year = {2012}
}

@INPROCEEDINGS{vmaf_paper,
  author={Lin, Joe Yuchieh and Liu, Tsung-Jung and Wu, Eddy Chi-Hao and Kuo, C.-C Jay},
  booktitle={Signal and Information Processing Association Annual Summit and Conference (APSIPA)}, 
  year ={2014},
  title={A fusion-based video quality assessment {(FVQA)} index}, 
  doi={10.1109/APSIPA.2014.7041705}}

@inproceedings{colmap,
    author={Sch\"{o}nberger, Johannes Lutz and Frahm, Jan-Michael},
    title={Structure-from-Motion Revisited},
    booktitle={Conference on Computer Vision and Pattern Recognition (CVPR)},
    year={2016},
}

@article{av1_stream_conf,
    author={{3GPP Technical Specification Group TSG SA}},
    title={{5G Video Codec Characteristics}},
    journal={TR 26.955, Release 17},
    year={2022}
}

@ARTICLE{technical_overview_av1,
  author={Han, Jingning and Li, Bohan and Mukherjee, Debargha and Chiang, Ching-Han and Grange, Adrian and Chen, Cheng and Su, Hui and Parker, Sarah and Deng, Sai and Joshi, Urvang and Chen, Yue and Wang, Yunqing and Wilkins, Paul and Xu, Yaowu and Bankoski, James},
  journal={Proceedings of the IEEE}, 
  title={A Technical Overview of AV1}, 
  year={2021},
  volume={109},
  number={9},
  pages={1435-1462},
  doi={10.1109/JPROC.2021.3058584}}

@INPROCEEDINGS{mov-slam,
  author={Turner, Richard N.C. and Banerjee, Natasha Kholgade and Banerjee, Sean},
  booktitle={2023 Seventh IEEE International Conference on Robotic Computing (IRC)}, 
  title={MoV-SLAM: Using Motion Vectors for Real-Time Single-CPU Visual SLAM}, 
  year={2023},
  volume={},
  number={},
  pages={51-58},
  doi={10.1109/IRC59093.2023.00015}}

@article{SIFT,
author = {Lowe, David G.},
title = {Distinctive Image Features from Scale-Invariant Keypoints},
year = {2004},
issue_date = {November 2004},
publisher = {Kluwer Academic Publishers},
address = {USA},
volume = {60},
number = {2},
issn = {0920-5691},
url = {https://doi.org/10.1023/B:VISI.0000029664.99615.94},
doi = {10.1023/B:VISI.0000029664.99615.94},
journal = {Int. J. Comput. Vision},
month = nov,
pages = {91–110},
numpages = {20}
}

@inproceedings{glomap,
  title={Global structure-from-motion revisited},
  author={Pan, Linfei and Bar{\'a}th, D{\'a}niel and Pollefeys, Marc and Sch{\"o}nberger, Johannes L},
  booktitle={European Conference on Computer Vision},
  pages={58--77},
  year={2024},
  organization={Springer}
}

@InProceedings{superpoint,
author = {DeTone, Daniel and Malisiewicz, Tomasz and Rabinovich, Andrew},
title = {SuperPoint: Self-Supervised Interest Point Detection and Description},
booktitle = {Proceedings of the IEEE Conference on Computer Vision and Pattern Recognition (CVPR) Workshops},
month = {June},
year = {2018}
}

@InProceedings{superglue,
author = {Sarlin, Paul-Edouard and DeTone, Daniel and Malisiewicz, Tomasz and Rabinovich, Andrew},
title = {SuperGlue: Learning Feature Matching With Graph Neural Networks},
booktitle = {Proceedings of the IEEE/CVF Conference on Computer Vision and Pattern Recognition (CVPR)},
month = {June},
year = {2020}
}

@article{disk,
  title={Disk: Learning local features with policy gradient},
  author={Tyszkiewicz, Micha{\l} and Fua, Pascal and Trulls, Eduard},
  journal={Advances in neural information processing systems},
  volume={33},
  pages={14254--14265},
  year={2020}
}

@inproceedings{lightglue,
  title={Lightglue: Local feature matching at light speed},
  author={Lindenberger, Philipp and Sarlin, Paul-Edouard and Pollefeys, Marc},
  booktitle={Proceedings of the IEEE/CVF international conference on computer vision},
  pages={17627--17638},
  year={2023}
}

@inproceedings{loftr,
  title={LoFTR: Detector-free local feature matching with transformers},
  author={Sun, Jiaming and Shen, Zehong and Wang, Yuang and Bao, Hujun and Zhou, Xiaowei},
  booktitle={Proceedings of the IEEE/CVF conference on computer vision and pattern recognition},
  pages={8922--8931},
  year={2021}
}

@article{h264-2d-3d,
  title={Conversion of H.264-encoded 2D video to 3D format},
  author={Mahsa T. Pourazad and Panos Nasiopoulos and Rabab Kreidieh Ward},
  journal={2010 Digest of Technical Papers International Conference on Consumer Electronics (ICCE)},
  year={2010},
  pages={63-64},
  url={https://api.semanticscholar.org/CorpusID:440814}
}

@article{orb_slam_3,
   title={ORB-SLAM3: An Accurate Open-Source Library for Visual, Visual–Inertial, and Multimap SLAM},
   volume={37},
   ISSN={1941-0468},
   url={http://dx.doi.org/10.1109/TRO.2021.3075644},
   DOI={10.1109/tro.2021.3075644},
   number={6},
   journal={IEEE Transactions on Robotics},
   publisher={Institute of Electrical and Electronics Engineers (IEEE)},
   author={Campos, Carlos and Elvira, Richard and Rodriguez, Juan J. Gomez and M. Montiel, Jose M. and D. Tardos, Juan},
   year={2021},
   month=dec, pages={1874–1890}
}

@INPROCEEDINGS{icir_2025_jz,
  author={Zouein, Julien and Javidnia, Hossein and Pitié, François and Kokaram, Anil},
  booktitle={2025 IEEE 4th International Conference on Intelligent Reality (ICIR)}, 
  title={Leveraging AV1 Motion Vectors for Fast and Dense Feature Matching}, 
  year={2025},
  volume={},
  number={},
  pages={1-4},
  doi={10.1109/ICIR68135.2025.11361611}}

@INPROCEEDINGS{pcs_2025_jz,
  author={Zouein, Julien and Vibhoothi, Vibhoothi and Kokaram, Anil},
  booktitle={2025 Picture Coding Symposium (PCS)}, 
  title={AV1 Motion Vector Fidelity and Application for Efficient Optical Flow}, 
  year={2025},
  volume={},
  number={},
  pages={1-5},
  doi={10.1109/PCS65673.2025.11417638}}

@article{nerf,
author = {Mildenhall, Ben and Srinivasan, Pratul P. and Tancik, Matthew and Barron, Jonathan T. and Ramamoorthi, Ravi and Ng, Ren},
title = {NeRF: representing scenes as neural radiance fields for view synthesis},
year = {2021},
issue_date = {January 2022},
publisher = {Association for Computing Machinery},
address = {New York, NY, USA},
volume = {65},
number = {1},
issn = {0001-0782},
url = {https://doi.org/10.1145/3503250},
doi = {10.1145/3503250},
journal = {Commun. ACM},
month = dec,
pages = {99–106},
numpages = {8}
}

@article{instant-ngp,
   title={Instant neural graphics primitives with a multiresolution hash encoding},
   volume={41},
   ISSN={1557-7368},
   url={http://dx.doi.org/10.1145/3528223.3530127},
   DOI={10.1145/3528223.3530127},
   number={4},
   journal={ACM Transactions on Graphics},
   publisher={Association for Computing Machinery (ACM)},
   author={Müller, Thomas and Evans, Alex and Schied, Christoph and Keller, Alexander},
   year={2022},
   month=jul, pages={1–15} }

@Article{3dgs,
      author       = {Kerbl, Bernhard and Kopanas, Georgios and Leimk{\"u}hler, Thomas and Drettakis, George},
      title        = {3D Gaussian Splatting for Real-Time Radiance Field Rendering},
      journal      = {ACM Transactions on Graphics},
      number       = {4},
      volume       = {42},
      month        = {July},
      year         = {2023},
      url          = {https://repo-sam.inria.fr/fungraph/3d-gaussian-splatting/}
}

@article{eap-gs,
  title={EAP-GS: Efficient Augmentation of Pointcloud for 3D Gaussian Splatting in Few-shot Scene Reconstruction},
  author={Dongrui Dai and Yuxiang Xing},
  journal={2025 IEEE/CVF Conference on Computer Vision and Pattern Recognition (CVPR)},
  year={2025},
  pages={16498-16507},
  url={https://api.semanticscholar.org/CorpusID:280089476}
}

@inproceedings{lpips,
  title={The unreasonable effectiveness of deep features as a perceptual metric},
  author={Zhang, Richard and Isola, Phillip and Efros, Alexei A and Shechtman, Eli and Wang, Oliver},
  booktitle={Proceedings of the IEEE conference on computer vision and pattern recognition},
  pages={586--595},
  year={2018}
}

@article{aomctc,
    title={{AOM Common Test Conditions v8.0}},
    author ={Zhao, X. and Lei, Z. (Ryan) and Norkin, A. and Daede, T. and Tourapis, A. and Vibhoothi,  V. and Pham, V. L. and Hoang, D. and Kuusela, A. and Golam Sarwer,  M.},
    Journal={Alliance for Open Media, Codec Working Group Output Document},
    volume = {CWG/F384o},
    year={2025},
    note={\url{https://aomedia.org/docs/CWG-F384o_AV2_CTC_v8.pdf}}   }

@Article{2023_gsplat_orig_paper,
      author       = {Kerbl, Bernhard and Kopanas, Georgios and Leimk{\"u}hler, Thomas and Drettakis, George},
      title        = {3D Gaussian Splatting for Real-Time Radiance Field Rendering},
      journal      = {ACM Transactions on Graphics},
      number       = {4},
      volume       = {42},
      month        = {July},
      year         = {2023},
      url          = {https://repo-sam.inria.fr/fungraph/3d-gaussian-splatting/}
}

@INPROCEEDINGS{2009_q_metric_paper_qomex,
  author={Zhu, Xiang and Milanfar, Peyman},
  booktitle={2009 International Workshop on Quality of Multimedia Experience}, 
  title={A no-reference sharpness metric sensitive to blur and noise}, 
  year={2009},
  volume={},
  number={},
  pages={64-69},
  doi={10.1109/QOMEX.2009.5246976}}

\end{document}